\documentclass[10pt,aps, prl,twocolumn,superscriptaddress]{revtex4-1}

\usepackage{graphicx}
\usepackage{hyperref}
\usepackage{amsmath}
\usepackage{bm}
\usepackage{xfrac}
\usepackage[all]{hypcap}
\hypersetup{colorlinks=true,linkcolor=blue,citecolor=blue,urlcolor=blue}

\usepackage{braket}
\usepackage{tikz}

\newcommand{\mytilde}{\raise.17ex\hbox{$\scriptstyle\mathtt{\sim}$}}
\newcommand{\UBC}{Department of Physics and Astronomy, University of British Columbia, Vancouver, Canada V6T 1Z1}
\newcommand{\QMI}{Quantum Matter Institute, University of British Columbia, Vancouver, Canada V6T 1Z4}
\newcommand{\DRES}{Max Planck Institute for Chemical Physics of Solids, N\"othnitzerstra\ss e 40, 01187 Dresden, Germany}

\begin{document}

\title{Bond disproportionation and dynamical charge fluctuations in the perovskite rare earth nickelates}

\author{R. J. Green}
\email[]{rgreen@phas.ubc.ca}
\affiliation{\UBC}
\affiliation{\QMI}
\affiliation{\DRES}

\author{M. W. Haverkort}
\affiliation{\DRES}

\author{G. A. Sawatzky}
\affiliation{\UBC}
\affiliation{\QMI}

%\date{\today}

\begin{abstract}
We present a theory describing the local electronic properties of the perovskite rare earth nickelates---materials which have negative charge transfer energies, strong O $2p$ -- Ni $3d$ covalence, and breathing mode lattice distortions at the origin of highly studied metal-insulator and antiferromagnetic ordering transitions. Utilizing a full orbital, full correlation double cluster approach, we find strong charge fluctuations in agreement with a bond disproportionation interpretation. The unique double cluster formulation permits the inclusion of necessary orbital degeneracies and Coulomb interactions to calculate resonant x-ray spectral responses, with which we find excellent agreement with well-established experimental results. This previously absent, crucial link between theory and experiment provides validation of the recently proposed bond disproportionation theory, and provides an analysis methodology for spectroscopic studies of engineered phases of nickelates and other high valence transition metal compounds.
\end{abstract}

\maketitle

\begin{figure*}
\includegraphics[width=7.0in]{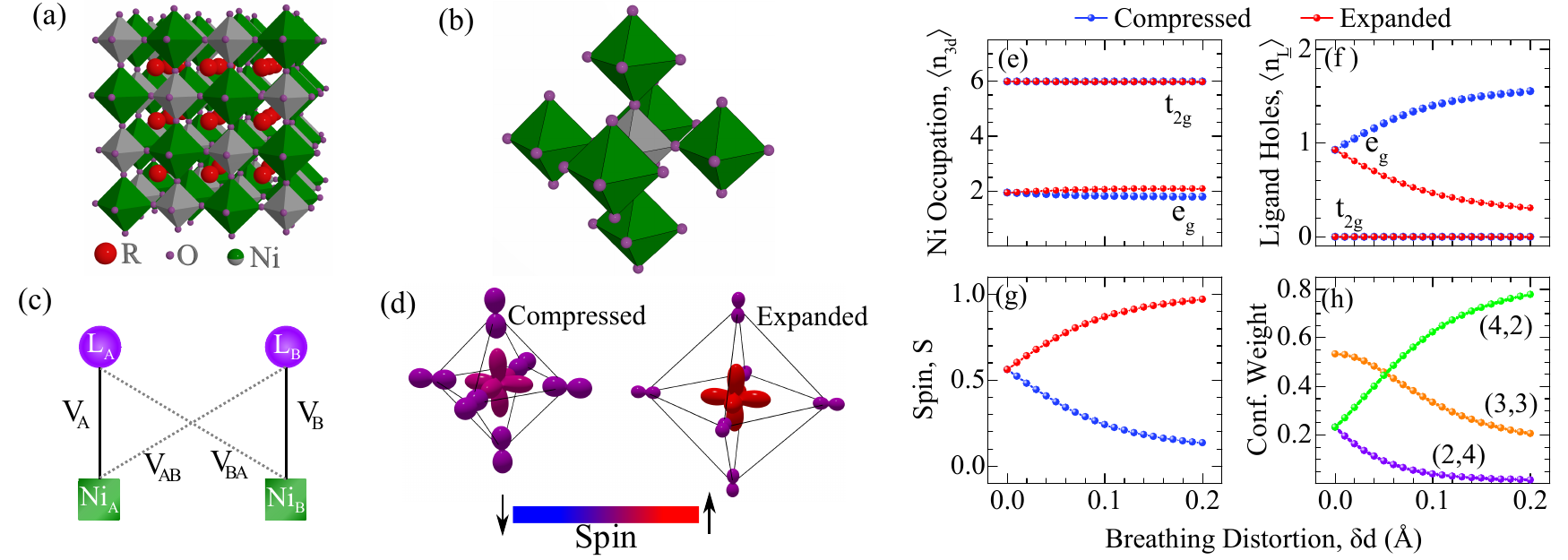}
\caption{(Color online) (a) Perovskite nickelate structure with alternating octahedra shaded to distinguish the long (green) and short (grey) bond sites in the low temperature distorted phase. (b) A cutout of the structure demonstrating the $O_h$ arrangment of the two octahedra types. (c) Diagrammatic depictions of the hopping arrangements for the double cluster model. (d) Visualization of calculated hole density matrices on compressed (left) and expanded (right) octahedra in the distorted phase (ligand hole densities are scaled by a factor of 2 for clarity). (e-h) Ground state characteristics from the double cluster model: (e) Distribution of Ni $3d$ electrons. (f) Distribution of ligand holes. (g) Spins for the two octahedra. (h) Weights of configurations with hole distributions ($n_A$,$n_B$) on the compressed and expanded octahedra, respectively.}
\label{Fig:1}
\end{figure*}

The perovskite rare-earth nickelate compounds ($\mathcal{R}$NiO$_3$) possess a rich phase diagram, exhibiting metal-insulator and antiferromagnetic ordering transitions with temperatures tunable via the rare earth ionic size \cite{Medarde_JPCM_1997}. The transition into the low temperature insulating phase is concomitant with a structural change where alternating NiO$_6$ octahedra are expanded and compressed in a rocksalt-pattern breathing mode distortion (see Fig. \ref{Fig:1}(a)). Below the N\'eel temperature in this distorted phase, the nickelates are E'-type antiferromagnets, with an uncommon $\mathbf{q}=\left(\sfrac{1}{4},\sfrac{1}{4},\sfrac{1}{4}\right)$ ordering vector.  Remarkably, pressure \cite{Medarde_JPCM_1997}, strain \cite{Catalano_SNO_APL_2014, Chak_StrainNNO_APL_2010, Frano_OrbControl_PRL_2013, Liu_NatCom_2013}, reduced dimensionality \cite{Boris_Science2011, Hepting_PRL2014}, and the engineering of various heterointerfaces \cite{Benckiser_OrbRefl_NatMat_2011, Disa_PRL2015, 2016MiddeyReview, Chen_PRL2013} have all been found to tune the ground state in various ways. Such a diversity of control mechanisms driving fascinating and useful emergent properties has fueled a wealth of interest in the nickelates in recent years.

A key property underlying the physics of the nickelates is the unusually high formal 3+ oxidation state imposed on the Ni ions. Oxides with Ni$^{3+}$ ($3d^7$) ions are relatively rare, and recent work suggests that the Ni here has a valence closer to 2+, with compensatory holes present in the oxygen $2p$ band.  Having such a negative charge transfer energy \cite{Mizokawa_NaCuO2_1991} in the ZSA classification scheme \cite{Zaanen_ZSA_PRL_1985}\footnote{Most studies (including Refs. \onlinecite{Mizokawa_NaCuO2_1991, Zaanen_ZSA_PRL_1985}, for example) define the charge transfer energy as the energy difference between the center of the oxygen \unexpanded{$2p$} band and the center of the upper Hubbard \unexpanded{$3d$} band. For small and negative charge transfer energies like the present work, the quantity of interest is the energy separation between the \emph{top} of the oxygen band and the \emph{bottom} of the \unexpanded{$3d$} band. Often called the effective charge transfer energy, this quantity can also be defined as the smallest energy difference between \unexpanded{$d^n$} and \unexpanded{$d^{n+1}\underline{L}$} configurations.}, the nickelates can accordingly be described as \emph{self-doped} Mott insulators \cite{Korotin_CrO2_PRL1998}. Theoretical studies which assume a negative charge transfer energy \emph{a priori} find a novel explanation for the metal-insulator transition in the form of bond disproportionation \cite{Mizokawa_SDMI_PRB_2000, 2007PRL_Mazin, ParkMillis_SSM_PRL_2012, Johnston_PRL_2014, LauMillis_PRL_2013, 2015_PRB_AGeorges}. Here, in the low temperature insulating phase, the self-doped oxygen $2p$ holes mix strongly with alternating Ni sites leading to a rocksalt-type superlattice distortion which can be identified roughly as collapsed $d^8\underline{L}^2$ and expanded $d^8\underline{L}^0$ Ni-O octahedra (where $\underline{L}$ denotes a ligand hole). A strong antiferromagnetic interaction between the ligand and Ni holes then leads reduced moments on the collapsed octahedra, such that short and long bond Ni have spins tending toward $S=0$ and $1$, respectively.

With indications of such a unique electronic structure at the origin of fascinating and tunable macroscopic phenomena, experiments which can directly probe the relevant properties of the nickelates are crucial. X-ray absorption spectroscopy (XAS) at the Ni $L_{2,3}$ edge has proven to be one such technique, having strong sensitivity to the metal-insulator transition (MIT) and negative charge transfer behavior \cite{PiamontezeXAS2002, Piamonteze_RNO_XAS_PRB_2005}, and clear trends exist in the XAS when moving across the phase diagram \cite{Freeland2015, Piamonteze_RNO_XAS_PRB_2005}. Closely related to the XAS is the resonant magnetic diffraction (RMD) response, which probes the antiferromagnetic ordering by studying the magnetic Bragg reflection with photon energies tuned to the Ni $L_{2,3}$ resonance. Recent studies have used XAS and RMD to characterize variations of the MIT due to strain \cite{Chak_StrainNNO_APL_2010, Meyers_Strain_PRB_2013}, to examine orbital polarization effects \cite{Benckiser_OrbRefl_NatMat_2011, Wu_PRB_OrbPol_2013}, and to study the collinearity of the magnetic moments \cite{Scagnoli2006, Frano_OrbControl_PRL_2013}. However, while the XAS and RMD show strong sensitivity to the electronic and magnetic structures, a satisfactory theoretical interpretation of each has yet to emerge for the nickelates. On one hand the bond disproportionation models studied to date do not include the necessary orbital degeneracies for core level spectroscopy analysis, and thus to date have not been verified experimentally. On the other hand the typical single cluster model approaches which are used for spectroscopy falter due to the negative charge transfer energy and breathing distortion. A bond disproportionation model which could be tested against XAS and RMD would simultaneously provide validation of the theory and a way to thoroughly analyze the ubiquitous, detailed spectra which have emerged from both bulk nickelates and engineered heterostructures \cite{Catalano_SNO_APL_2014, Chak_StrainNNO_APL_2010, Frano_OrbControl_PRL_2013, Benckiser_OrbRefl_NatMat_2011, Disa_PRL2015, Meyers_Strain_PRB_2013, Wu_PRB_OrbPol_2013, 2016MiddeyReview, Scagnoli2006, PiamontezeXAS2002, Piamonteze_RNO_XAS_PRB_2005, Freeland2015}.

In this work, we formulate a novel full correlation, double cluster model to describe the insulating and magnetic phases of the nickelates. The formulation permits the inclusion of negative charge transfer, bond- and charge-disproportionation and, most importantly, the necessary orbital degeneracies and Coulomb interactions to simulate core level spectroscopy, providing a key test of bond disproportionation against resonant x-ray experiments. Our approach finds the same instability toward $d^8\underline{L}^2$ ($S=0$) and $d^8\underline{L}^0$ ($S=1$) alternating octahedra in the presence of a breathing distortion that was found in recent restricted-orbital studies. Further, we find moment sizes for each sublattice in agreement with those determined experimentally. At the same time, we find spectral responses for XAS and RMD which are in excellent agreement with experiment, including pronounced trends across the rare earth series. While the model provides a previously missing validation of bond disproportionation against experiment, it is also a powerful tool for the precise analysis of engineered phases of nickelates as well as many other high valence transition metal oxides. 

\begin{figure*}
\includegraphics[width=7in]{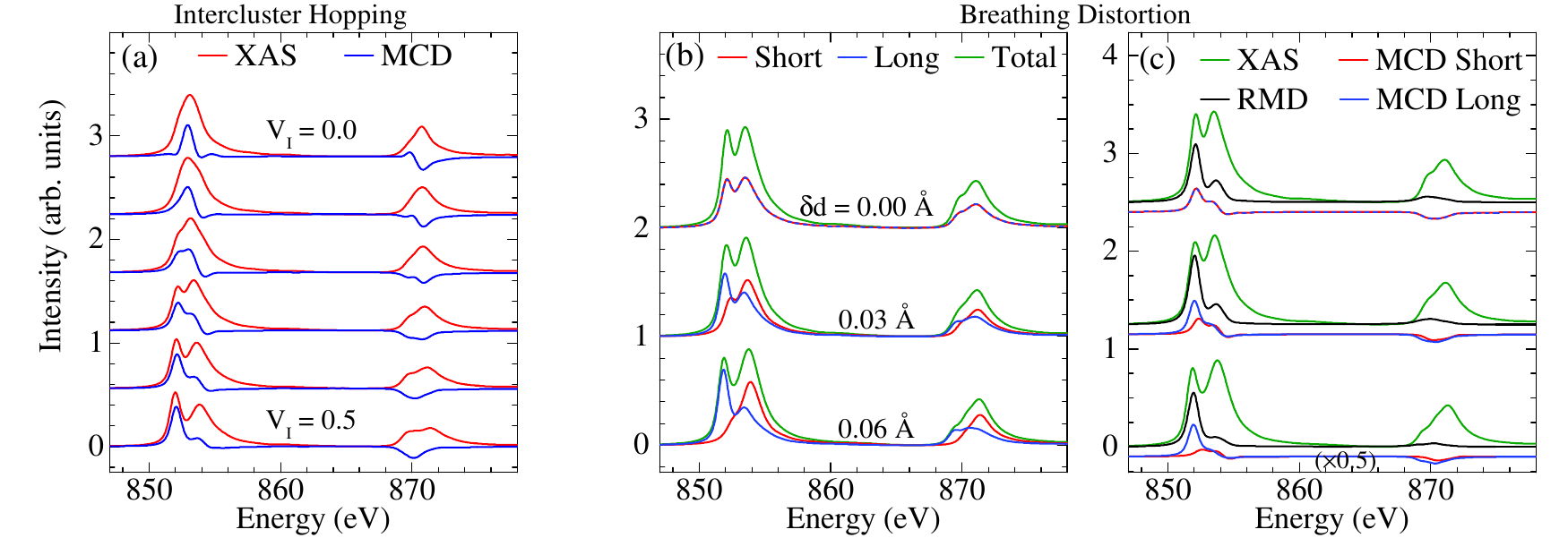}
\caption{(Color online) Resonant x-ray responses of the double cluster model. (a) XAS and MCD spectra are shown in the single cluster limit (top, intercluster hopping $V_{I}$ set to zero), and for increasing intercluster hopping values. (b) XAS spectra for different magnitudes of the breathing distortion (with $V_I = 0.35$). The spectra of the two inequivalent sites are shown, along with their sum. (c) Resonant magnetic diffraction and site-decomposed MCD spectra are shown for the same breathing distortions as (b), along with the total XAS for comparison. MCD spectra are offset for clarity.}
\label{Fig:Spec}
\end{figure*}

In Fig. \ref{Fig:1}(a), we depict the prototypical structure of the nickelates. Alternating octahedra are shaded to distinguish the long (green) and short (grey) bond NiO$_6$ octahedra present in the breathing mode distorted insulating phase. This distortion forms a rocksalt pattern, with each long (short) bond octahedron surrounded by six short (long) bond octahedra in an $O_h$ symmetric manner as exemplified in Fig. \ref{Fig:1}(b). To enable a calculation which includes the full local Coulomb interactions necessary for core level spectroscopy simulations, we develop a model similar in nature to the multiplet ligand field theory approximation which has had great success in this area \cite{Haverkort_Wannier_PRB_2010, degroot2008}. However, instead of the typical approach of a exact diagonalization on a single NiO$_{6}$ cluster, we create two clusters, one representing each sublattice of the rocksalt distortion. Each of our two Ni-O octahedral clusters are constructed from a standard ligand field theory Hamiltonian, including all local $3d$ (and Ni $2p$) Coulomb interactions, crystal field splittings, $3d$-ligand hybridization, and spin-orbit interaction \cite{SupMat_Green}. Recalling the rocksalt pattern formed by the two distinct octahedra in the solid, we then couple our two cluster models via $O_h$ symmetric hybridization operators (having $e_g$ and $t_{2g}$ symmetry). The coupling is shown diagrammatically in Fig. \ref{Fig:1}(c). The individual clusters have standard ligand field Ni-ligand hopping $V_A$ and $V_B$, and intercluster Ni-ligand hopping is shown as $V_{AB}$ and $V_{BA}$. The general form of our Hamiltonian is then $H = H_{LF_A} + H_{LF_B} + H_{mix}$ where the first two terms are the independent ligand field Hamiltonians for the two distinct octahedra, and the final term introduces the $O_h$-symmetric mixing (full details of the Hamiltonian and parameters are presented in the Supplemental Material \cite{SupMat_Green}). The breathing distortion is introduced into the model by a modulation of the hopping and crystal field terms, according to Harrison's rules \cite{Harrison1983, LauMillis_PRL_2013, Johnston_PRL_2014, SupMat_Green}. The computations are performed with our second quantization programming language and exact diagonalization code, \emph{Quanty} \cite{Haverkort_Wannier_PRB_2010, Haverkort_DMFT_2014, Haverkort_DMFTXAS_EPL2014, QuantyWeb}. 

Figures \ref{Fig:1}(d-h) display the ground state properties of the model as function of the breathing distortion, $\delta d$. We parameterize the breathing distortion as the difference in long and short bond lengths $d_L$ and $d_S$ from the mean value $d_0$ ($d_{\sfrac{L}{S}}=d_0~ \sfrac{+}{-} ~\delta d$). In Fig. \ref{Fig:1}(e), we find very little change in the Ni $3d$ occupation occurs via the breathing distortion. The $t_{2g}$ orbitals stay fully occupied and only a minor variation of the $e_g$ occupation occurs, with a total average occupation of \mytilde 8 electrons per Ni, consistent with the negative charge transfer scenario \footnote{We use a configuration-averaged charge transfer energy of \unexpanded{$\Delta = -0.5$} eV, which yields an effective charge transfer energy between the lowest \unexpanded{$d^7$} and \unexpanded{$d^8\underline{L}$} configurations of about \unexpanded{$-0.6$} eV \cite{SupMat_Green}}. 

The oxygen orbitals, in contrast to the Ni, are very active under the breathing distortion.  As shown in Fig. \ref{Fig:1}(f), with no distortion there is one (self-doped) $e_g$ ligand hole per cluster. However, the holes shift to the compressed octahedron when the breathing distortion is introduced. This action of the oxygen holes leads to a reduction of the spin on the compressed octahedron, as shown in Fig. \ref{Fig:1}(g), since the holes are strongly antiferromagnetically coupled to the holes in the Ni $3d$ shell. This leaves the expanded octahedron Ni unscreened, and its spin accordingly approaches $S=1$, the expected value for a high spin, ionic $3d^8$ configuration. For HoNiO$_3$, where breathing distortions and spin moments have been measured experimentally \cite{HoNiO3_Moments2001, HoNiO3_BondLengths2000}, we find good agreement with our model. The general behavior of Figs. \ref{Fig:1}(e-g), and summarized pictorially for $\delta d=0.05$\AA~in Fig. \ref{Fig:1}(d), also agrees with recent theory studies which were restricted to the active $e_g$ orbital basis \cite{ParkMillis_SSM_PRL_2012, LauMillis_PRL_2013, Johnston_PRL_2014}.

Overall, the calculations show a clear transition from a $\left(3,3\right)$ hole occupation of the two octahedra toward a $\left(4,2\right)$ hole arrangement for the compressed and expanded octahedra, respectively, with spins accordingly transitioning from $\left(\sfrac{1}{2},\sfrac{1}{2}\right)$ to $\left(0,1\right)$. This transition is detailed in Fig. \ref{Fig:1}(h), where we plot the projection of the ground state onto the relevant basis states. We emphasize that even with no breathing distortion a dynamic charge ordering is present, while the bond disproportionated configuration $\left(4,2\right)$ dominates the ground state for large distortions.

In Figure \ref{Fig:Spec} we show the resonant soft x-ray responses of our model, which has the full Ni $3d$ and $2p$ orbital degeneracies and their Coulomb interactions necessarily included in order to capture the detailed multiplet features in the spectra. In Fig. \ref{Fig:Spec}(a), we compare the double cluster model (with no breathing distortion) to a conventional single cluster model, where the intercluster hopping $V_I$ is set to zero. Increasing $V_I$ from zero leads to a pronounced first peak being separated out below both the $L_3$ and $L_2$ edges. Such a first $L_3$ peak is a distinct characteristic of the insulating nickelates \cite{Freeland2015} and, importantly, we find that the peak also has a strong magnetic circular dichroism (MCD) signal, consistent with RMD experiments which find the magnetic diffraction to be strongest at that peak \cite{Scagnoli2006, Bodenthin_JPCM_2011, Frano_OrbControl_PRL_2013}. Given that no breathing distortion is introduced yet in this plot, the peak arises due to non-local excitations between the two clusters. Such an effect is perhaps not surprising, given the negative charge transfer energy and strong covalence, but it demonstrates that even in the absence of breathing distortions, a proper interpretation of the spectra of high valence transition metal oxides might require at least a double cluster model where such excitations can be captured. Similar inter-site effects requiring extensions beyond the single cluster approximation were recently identified in the XAS of strongly covalent mixed valence systems \cite{Gupta_InterXAS_EPL2011}. We note that our value of $V_I=0.35$ is slightly less than one might expect for 180 degree bonds between adjacent Ni atoms \cite{SupMat_Green}, in accordance with octahedral tilting that is present in the real materials.

Figure \ref{Fig:Spec}(b) shows the effect of the breathing distortion on the XAS spectra. In the presence of the distortion there are distinct spectra, one arising from each of the two inequivalent sites, and as such we show these two spectra, along with their sum. Here it is evident that the spectrum from the long (short) bond octahedron shifts to lower (higher) energies as the breathing distortion is increased. In the extreme case, the sharp pre-peak arises entirely from the long bond Ni. This trend is in excellent agreement with the nickelate phase diagram, where smaller rare earths lead to larger breathing distortions and a larger separation of the XAS peaks \cite{Freeland2015, Piamonteze_RNO_XAS_PRB_2005}.

Figure \ref{Fig:Spec}(c) details the magnetic response for the same three breathing distortions of Fig. \ref{Fig:Spec}(b).  The total XAS is shown again for comparison, and the site-decomposed MCD spectra are shown. Consistent with the tendency toward $S=0$ and $S=1$ moments on the short and long bond sites, as was shown in Fig. \ref{Fig:1}, the MCD spectrum of the long (short) bond site becomes stronger (weaker) as the breathing distortion is introduced. As the long bond site contributes the most to the sharp first peak, the MCD response is then concentrated mostly on that peak as well. For comparison to experiment, we also show the calculated resonant magnetic diffraction (RMD) response, which peaks strongly at the XAS first peak and has a weak shoulder at higher energies, in excellent agreement with previous experiments \cite{Scagnoli2006, Bodenthin_JPCM_2011, Frano_OrbControl_PRL_2013}.

\begin{figure}
\includegraphics[width=3.375in]{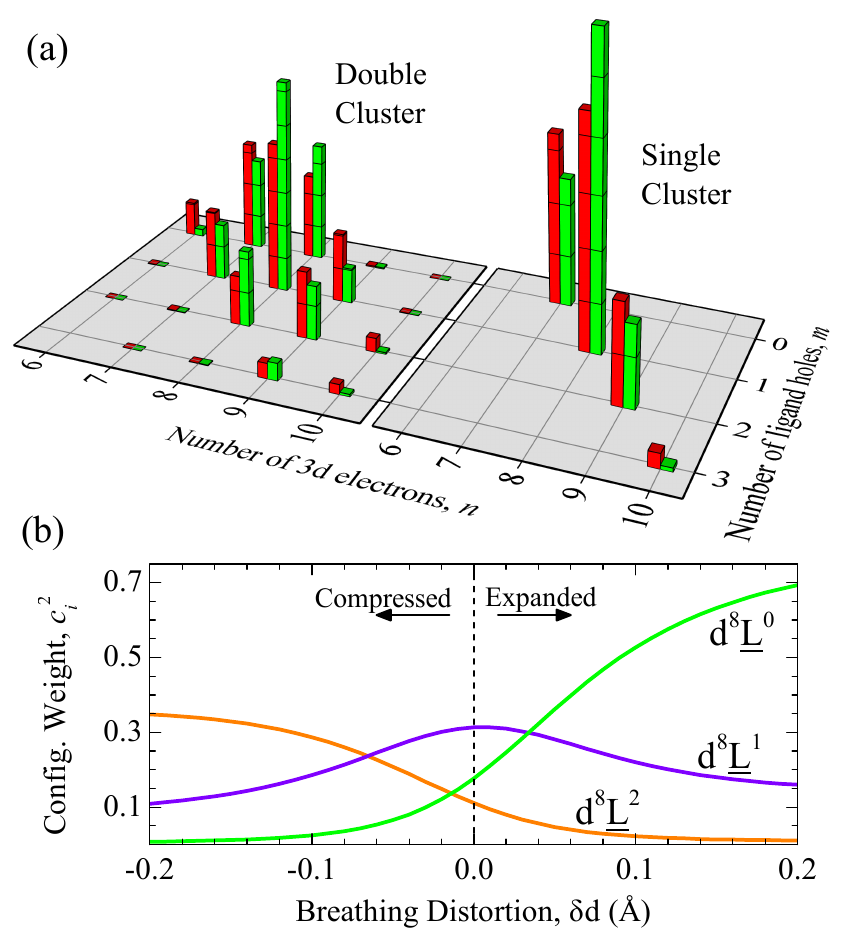}
\caption{(Color online) (a) Configuration weights $c_i^2$ are compared for the present double cluster model (left, green) and conventional ligand field theory (right, green), demonstrating the new determinants which arise in negative charge transfer situations not captured by single cluster theory. In both cases the weights for the non-interacting approximation are shown in red for comparison. Divisions on the bars have values of 0.05 for the double cluster and 0.10 for the single. (b) The compressed octahedron exhibits a wavefunction with dominant ligand hole-rich $d^8$ determinants, while the expanded octahedron has a much more ionic wavefunction. }
\label{Fig:Char}
\end{figure}

The effects of intercluster hopping on the XAS spectra of Fig. \ref{Fig:Spec}(a) demonstrate the importance of non-local excitations and accordingly suggest a highly covalent nature of the ground state captured in our double cluster model. We verify this in Fig. \ref{Fig:Char} by a decompositional analysis of the ground state wavefunction. Within the configuration interaction approach taken here, the wavefunction is built up of determinants belonging to specific electronic configurations. In conventional single cluster theory, the ground state is thus given by $|\psi_0\rangle = \sum _i c_i |d^{n+i}\underline{L}^i\rangle$, where $c_i^2$ gives the weight of configuration $i$. In the nickelates, $n=7$, and one is therefore limited to 4 configurations. The weight of these configurations in the single cluster limit of our model are plotted on the right in Fig. \ref{Fig:Char}(a).  In green we show the calculations including electron-electron correlations, i.e. the Hamiltonian including the full Coulomb interaction is solved by exact diagonalization. In red we show the weight of the different configurations approximating the Coulomb interaction by a potential as one would do in a local density or Hartree-Fock approximation. Correlations induced by the Coulomb interaction reduce the amount of charge fluctuations. One configuration gains more weight, reducing the weight of all others and thus changing the variance of the Ni $d$ electron count. Interestingly it is not the $|d^{7}\underline{L}^0\rangle$ configuration that gains the most weight in the strongly correlated limit, but the $|d^{8}\underline{L}^1\rangle$ configuration. The local symmetry is still that of a $|d^{7}\underline{L}^0\rangle$ configuration, i.g. $\langle S^2 \rangle =3/4$ and one has a local doublet, defining the formal valence to be Ni$^{3+}$.

In the double cluster model many more configurations are possible ($d^n\underline{L}^m$, in general) due to charge fluctuations between the clusters. We plot the configuration weights for the double cluster model with no breathing distortion on the left in Fig. \ref{Fig:Char}(a). Again correlations (green) reduce the amount of charge fluctuations on the Ni site and, like the single cluster calculation, the $d^8$ configurations have the most weight. The ligand charge fluctuations are not reduced between the correlated (green) and uncorrelated (red) calculations. The local state can thus be thought of as a superposition of $|d^{8}\underline{L}^0\rangle$, $|d^{8}\underline{L}^1\rangle$, $|d^{8}\underline{L}^2\rangle$. Magnetically the local cluster state is thus a superposition of a triplet, doublet and singlet with Ni-ligand cluster formal valences of 2+, 3+, and 4+ respectively.

In Fig. \ref{Fig:Char}(b), we show the effects of the breathing distortion on the $d^8$ derived configurations of the double cluster model. As demonstrated above, for no breathing distortion there is a strong mixture of configurations (high covalency). However, upon the introduction of the breathing distortion the ligand hole character strongly shifts, such that the compressed octahedron has mostly $d^8\underline{L}^2$ (and to a lesser extent, $d^8\underline{L}^1$) character (reflecting what was shown in Fig. \ref{Fig:1}(f)). The expanded octahedron, however, becomes much more ionic, being dominated by the $d^8\underline{L}^0$ configuration. It is this ionic character of the long bond sublattice which has led to suggestions that the nickelate MIT can be described as a \emph{site selective} Mott transition \cite{ParkMillis_SSM_PRL_2012}. Alternatively, given the active oxygen electronic structure due to the negative charge transfer energy, and the lattice breathing distortion, the MIT might be more aptly described as Peierls-like.

Our double cluster model bridges the gap between electronic structure experiments and theories of negative charge transfer and bond disproportionation in the rare earth nickelates. By accounting for self-doped holes in the oxygen $2p$ band as well as structural breathing distortions, we find pronounced bond disproportionation effects in the ground state, including unequal spin moments on the two sublattices in excellent agreement with experiments. Additionally, we show the effect of bond disproportionation on the soft x-ray spectral response, again finding excellent agreement with experiment and providing a tool for future studies of engineered phases. The highly covalent wavefunctions and non-local excitations captured by the double cluster model developed here show that it will be important for other high oxidation state transition metal oxides, even in the absence of breathing distortions. New insights into highly studied materials such as perovskite manganates, ferrates, cobaltates, etc., and their resonant scattering responses, should also be captured by the model.

\begin{acknowledgments}
We thank M. Hepting, E. Benckiser, Y. Lu, I. Elfimov, M. Berciu, T. Schmitt, and V. Bisogni for helpful discussions. This work was supported by NSERC, CIfAR, and the Max Planck - UBC Centre for Quantum Materials.
\end{acknowledgments}

%

%temp packages
\newcommand\FramedBox[3]{%
  \setlength\fboxsep{0pt}
  \fbox{\parbox[t][#1][c]{#2}{\centering\huge #3}}}

\newcommand\SUpBig[1][2ex]{\mathrel{\rotatebox{90}{$\xrightarrow{\rule{#1}{0pt}}$}}}

\NewDocumentCommand\SpinDn{O{2.0ex} O{black}}{%
   \mathrel{\tikz[baseline] \draw [<-, line width=0.5pt, #2] (0,0) -- ++(0,#1);}}
\NewDocumentCommand\SpinUp{O{2.0ex} O{black}}{%
   \mathrel{\tikz[baseline] \draw [->, line width=0.5pt, #2] (0,0) -- ++(0,#1);}}

\renewcommand*{\citenumfont}[1]{S#1}
\renewcommand*{\bibnumfmt}[1]{[S#1]}

\makeatletter 
\renewcommand{\thefigure}{S\@arabic\c@figure}
\makeatother

\makeatletter 
\renewcommand{\thetable}{S\@arabic\c@table}
\makeatother

\makeatletter 
\renewcommand{\theequation}{S\@arabic\c@equation}
\makeatother

\setcounter{equation}{0}
\setcounter{figure}{0}
\setcounter{table}{0}
\setcounter{page}{0}

\widetext
\clearpage

\begin{center}
\textbf{\emph{Supplementary Information for} ``Bond disproportionation and dynamical charge fluctuations in the perovskite rare earth nickelates''}
\end{center}

\section{Model Hamiltonian}

\subsection*{Main Terms}

Our model consists of two Ni-Ligand clusters, roughly equivalent to two NiO$_6$ octahedra, each containing a Ni $2p$ shell, a Ni $3d$ shell, and a ligand shell. The ligand shell orbitals are defined as linear combinations of the actual oxygen $2p$ like Wannier-orbitals having the same rotation properties as the Ni $d$ orbitals within the $O_h$ point group used here \cite{Ballhausen1968S,Haverkort_Wannier_PRB_2010S}. The general form of our Hamiltonian is:
\begin{equation}
H = H_{LF_A} + H_{LF_B} + H_{mix},
\end{equation}
where $A$ and $B$ refer to the short and long bond sites, respectively (or identical sites when no breathing distortion is present) and $H_{mix}$ is the part of the Hamiltonian that couples the two clusters together. Taken separately, $H_{LF_A}$ and $H_{LF_B}$ are independent multiplet ligand field theory Hamiltonians. For $H_{LF_A}$, we have:
\begin{equation}
H_{LF_A} = H_U^{dd} + H_U^{pd} + H_{\bm{l} \cdot \bm{s}}^{d} + H_{\bm{l} \cdot \bm{s}}^{p} + H_{o}^{p} + H_{o}^{d} + H_{o}^{L} + H_{hyb}^{dL},
\end{equation}
with, \\
\begin{tabular}{ll}
$H_U^{dd}$ & the Coulomb repulsion between two Ni $3d$ electrons including all multiplet effects,\\
$H_U^{pd}$ & the Coulomb repulsion between a Ni $2p$ core and $3d$ valence electron including all multiplet effects,\\
$H_{\bm{l} \cdot \bm{s}}^{d}$ & the Ni $3d$ spin orbit interaction,\\
$H_{\bm{l} \cdot \bm{s}}^{p}$ & the Ni $2p$ core level spin orbit interaction,\\
$H_{o}^{p}$ & the onsite energy of the Ni $2p$ core orbitals,\\
$H_{o}^{d}$ & the orbital dependent onsite energy of the Ni $3d$ valence orbitals,\\
$H_{o}^{L}$ & the orbital dependent onsite energy of the Ligand orbitals, and\\
$H_{hyb}^{dL}$ & the hybridization strength between the Ni $3d$ and Ligand orbitals.
\end{tabular}

The Hamiltonian for site $B$ is analogous to $H_{LF_A}$. Below we list each term in the Hamiltonian in more detail.

\subsection*{\texorpdfstring{On-site Energy of the Ni $2p$ Orbitals - $H_{o}^{p}$}{}}

The onsite energy of the Ni $2p$ core electrons is given as:
\begin{equation}
H_{o}^{p} = \epsilon_p \sum_{\tau} \bm{p}^{\dag}_{\tau} \bm{p}^{\phantom{\dag}}_{\tau},
\end{equation}
with $\tau$ labeling the 6 different Ni $2p$ spin-orbitals with $m=-1,0,1$ and $\sigma=\pm1/2$, $\bm{p}^{\dag}_{\tau}$ ($\bm{p}^{\phantom{\dag}}_{\tau}$) the operator creating (annihilating) an electron in orbital $\tau$ and $\epsilon_p$ defined in terms of $U_{dd}$, $U_{pd}$ and $\Delta$ as \cite{Zaanen_ZSA_PRL_1985S, Zaanen_NiPES_PRB_1986S, degroot2008S}:
\begin{equation}
\epsilon_{p} = \frac{ 10\Delta + \left(1+n_d\right) \left(n_d\frac{U_{dd}}{2}-\left(10+n_d\right)U_{pd}\right)}{16+n_d}
\end{equation}
where $n_d$ is the formal number of $3d$ electrons per Ni ($n_d = 7$ for the nickelates studied here).

\subsection*{\texorpdfstring{On-site Energy of the Ni $3d$ Orbitals - $H_{o}^{d}$}{}}

The onsite energy of the Ni $3d$ valence electrons is given as:
\begin{equation}
H_{o}^{d} = \epsilon_d \sum_{\tau} \bm{d}^{\dag}_{\tau} \bm{d}^{\phantom{\dag}}_{\tau} + Dq_i \left( 6 \sum_{\tau\in e_g} \bm{d}^{\dag}_{\tau} \bm{d}^{\phantom{\dag}}_{\tau} - 4 \sum_{\tau\in t_{2g}} \bm{d}^{\dag}_{\tau} \bm{d}^{\phantom{\dag}}_{\tau} \right),
\end{equation}
with $\tau$ labeling the 10 different Ni $3d$ spin-orbitals belonging either to the $t_{2g}$ irreducible representation ($yz$, $xz$, and $xy$) or to the $e_g$ irreducible representation ($3z^2-r^2$, and $x^2-y^2$) with either spin up or spin down, $\bm{d}^{\dag}_{\tau}$ ($\bm{d}^{\phantom{\dag}}_{\tau}$) the operator creating (annihilating) an electron in orbital $\tau$, $\epsilon_d$ the shell average energy defined in terms of $U_{dd}$, $U_{pd}$ and $\Delta$ as \cite{Zaanen_ZSA_PRL_1985S, Zaanen_NiPES_PRB_1986S, degroot2008S}:
\begin{align}
\epsilon_d = \frac{10\Delta - n_d\left(31+n_d\right)\frac{U_{dd}}{2}-90U_{pd}}{16+n_d},
\end{align}
and $Dq_i$ the onsite part of the cubic crystal-field splitting. The value of $Dq_i$ depends on the the breathing distortion, being larger for the smaller site. In terms of the non-breathing value $Dq_0$, the breathing distortion is approximated by Harrison's rules for hybridization \cite{Harrison1983S} as
\begin{align}
Dq_i = Dq_0 \left(1 + \frac{\delta d_i}{d_0}\right)^{-4}
\end{align}
where, $d_0$ is the average bond length and $\delta d$ is the positive (negative) displacement from average for the long (short) bond octahedron $B$ ($A$).

\subsection*{\texorpdfstring{On-site Energy of the Ligand Orbitals - $H_{o}^{L}$}{}}

The ligand orbitals are linear combinations of the valence states of the infinite solid such that the local Ni $d$ orbitals directly interact only with these orbitals \cite{Haverkort_Wannier_PRB_2010S}. For each Ni spin-orbital there is exactly one Ligand orbital, independent of crystal symmetry and breathing distortion. The ligand orbitals can have a different onsite energy depending if they belong to the local $t_{2g}$ or $e_g$ irreducible representation. The Hamiltonian thus takes a very similar form as $H_{o}^{d}$ namely:
\begin{equation}
H_{o}^{L} = \epsilon_L \sum_{\tau} \bm{L}^{\dag}_{\tau} \bm{L}^{\phantom{\dag}}_{\tau} +  Tpp_i \left( \sum_{\tau\in e_g} \bm{L}^{\dag}_{\tau} \bm{L}^{\phantom{\dag}}_{\tau} - \sum_{\tau\in t_{2g}} \bm{L}^{\dag}_{\tau} \bm{L}^{\phantom{\dag}}_{\tau} \right),
\end{equation}
with $\tau$ labeling the 10 different Ligand spin-orbitals belonging either to the $t_{2g}$ irreducible representation ($yz$, $xz$, and $xy$) or to the $e_g$ irreducible representation ($3z^2-r^2$, and $x^2-y^2$) with either spin up or spin down, $\bm{L}^{\dag}_{\tau}$ ($\bm{L}^{\phantom{\dag}}_{\tau}$) the operator creating (annihilating) an electron in orbital $\tau$, $\epsilon_L$ the shell average energy defined in terms of $U_{dd}$, $U_{pd}$ and $\Delta$ as \cite{Zaanen_ZSA_PRL_1985S, Zaanen_NiPES_PRB_1986S, degroot2008S}:
\begin{align}
\epsilon_{L} = \frac{\left(1+n_d\right)\left(n_d\frac{U_{dd}}{2}+6U_{pd}\right) - \left(6+n_d\right)\Delta}{16+n_d},
\end{align}
and $T_{pp,i}$ roughly the hopping strength between two ligand O $2p$ orbitals \cite{Haverkort_Wannier_PRB_2010S}. The value of $T_{pp,i}$ depends on the breathing distortion and can be expressed in terms of $\delta d_i$, the positive (negative) displacement from average for the long (short) bond octahedron $B$ ($A$) and the non-breathing value $T_{pp}$, using rules defined by Harrison \cite{Harrison1983S} as
\begin{align}
T_{pp,i} = T_{pp} \left(1 + \frac{\delta d_i}{d_0}\right)^{-3}.
\end{align}

\subsection*{\texorpdfstring{Hybridization Between Ni $3d$ and Ligand Orbitals - $H_{hyb}^{dL}$}{}}

The interaction between the Ni $3d$ orbitals and the Ligand orbitals is given as:
\begin{equation}
\label{Eqn:Hhyb}
H_{hyb}^{dL} =  \sqrt{1-x} \left( V_{e_{g}} \sum_{\tau\in e_g} \left( \bm{d}^{\dag}_{\tau} \bm{L}^{\phantom{\dag}}_{\tau} + \bm{L}^{\dag}_{\tau} \bm{d}^{\phantom{\dag}}_{\tau} \right) + V_{t_{2g}} \sum_{\tau\in t_{2g}} \left( \bm{d}^{\dag}_{\tau} \bm{L}^{\phantom{\dag}}_{\tau} + \bm{L}^{\dag}_{\tau} \bm{d}^{\phantom{\dag}}_{\tau} \right) \right),
\end{equation}
with $\tau$ labeling the 10 different Ni $3d$ or Ligand spin-orbitals belonging either to the $t_{2g}$ irreducible representation ($yz$, $xz$, and $xy$) or to the $e_g$ irreducible representation ($3z^2-r^2$, and $x^2-y^2$) with either spin up or spin down, $\bm{d}^{\dag}_{\tau}$, $\bm{L}^{\dag}_{\tau}$ ($\bm{d}^{\phantom{\dag}}_{\tau}$, $\bm{L}^{\phantom{\dag}}_{\tau}$) the operator creating (annihilating) an electron in orbital $\tau$ and either the $d$ or Ligand shell. $V_{e_{g}}$ and $V_{t_{2g}}$ are the individual hopping strengths between the $d$ and Ligand orbitals. The parameter $x$ determines the ratio between the hopping within a single ligand field cluster and between two ligand-field clusters.

\subsection*{\texorpdfstring{Coupling Between Cluster $A$ and $B$ - $H_{mix}$}{}}

The interaction between cluster $A$ and cluster $B$ is given as:
\begin{align}
\label{Eqn:Hmix}
H_{mix} =  \sqrt{x} \bigg( &V_{e_{g}} \sum_{\tau\in e_g} \left( \bm{d}^{\dag}_{A,\tau} \bm{L}^{\phantom{\dag}}_{B,\tau} + \bm{L}^{\dag}_{B,\tau} \bm{d}^{\phantom{\dag}}_{A,\tau} + \bm{d}^{\dag}_{B,\tau} \bm{L}^{\phantom{\dag}}_{A,\tau} + \bm{L}^{\dag}_{A,\tau} \bm{d}^{\phantom{\dag}}_{B,\tau} \right) \\ 
\nonumber &+  V_{t_{2g}} \sum_{\tau\in t_{2g}} \left( \bm{d}^{\dag}_{A,\tau} \bm{L}^{\phantom{\dag}}_{B,\tau} + \bm{L}^{\dag}_{B,\tau} \bm{d}^{\phantom{\dag}}_{A,\tau} + \bm{d}^{\dag}_{B,\tau} \bm{L}^{\phantom{\dag}}_{A,\tau} + \bm{L}^{\dag}_{A,\tau} \bm{d}^{\phantom{\dag}}_{B,\tau}\right) \bigg),
\end{align}
with the individual terms defined as in the previous section. The definition of the hybridization interaction using the parameter $x$ is such that the overall hopping strength is independent of the coupling between the two clusters. For perfect periodic boundary conditions $x=\frac{1}{2}$. In this case one can create  bonding and anti-bonding linear combinations of the ligand orbitals of cluster $A$ and cluster $B$ and the anti-bonding linear combination will be non-bonding with respect to the Ni $3d$ orbitals. Although this senario on first sight looks like a good cluster model, it highly overestimates the Ni-Ni exchange interactions. For nearest neighbor clusters only one out of six oxygens is shared and one might expect $x\approx 1/6$ to yield reasonable results, assuming 180 degree Ni-O-Ni bonds. We used $x$ as an empirical parameter, as one can not claim convergence with respect to cluster size for a two site calculation. Due to the presence of non-local excitations, the XAS spectrum is quite sensitive to $x$ and we find best agreement with experiment using $x = 0.35^2 = 0.1225$, which is quite close to the expected value of $1/6$. The fact that we find a value slightly smaller than $1/6$ is likely due in part to the octahedral tilts in the real materials, which decrease the Ni-O-Ni bond angle from 180 degrees and reduce the effective Ni-Ni hopping. 

\subsection*{\texorpdfstring{Coulomb Repulsion Between Two Ni $3d$ Electrons - $H_{U}^{dd}$}{}}

The onsite Coulomb repulsion between two $d$ electrons is defined as:
\begin{align}
H_{U}^{dd} &= \sum_{i,j} \frac{1}{2}\frac{e^2}{|r_i-r_j|}\\
\nonumber  &= \sum_{\tau_1,\tau_2,\tau_3,\tau_4} U_{\tau_1,\tau_2,\tau_3,\tau_4} \bm{d}^{\dag}_{\tau_1} \bm{d}^{\dag}_{\tau_2} \bm{d}^{\phantom{\dag}}_{\tau_3} \bm{d}^{\phantom{\dag}}_{\tau_4},
\end{align}
with,
\begin{align}
 U_{\tau_1,\tau_2,\tau_3,\tau_4} = -\frac{1}{2} \delta_{\sigma_1,\sigma_3} \delta_{\sigma_2,\sigma_4}
 \sum_{k=0,2,4} 
 c^{\left(k\right)}\left[l_1=2,m_1;l_3=2,m_3\right]
 c^{\left(k\right)}\left[l_4=2,m_4;l_2=2,m_2\right]
 \times F^{\left(k\right)},
\end{align}
where $\tau$ are combined spin and orbital indices, $\sigma$ are spin indices, $l$ and $m$ are angular momentum indices ($l=2$ for $d$ electrons), $F^{(k)}$ are the radial (Slater) integrals, and 
\begin{align}
c^{\left(k\right)} \left[l_1,m_1;l_2,m_2\right] = 
\Bra{Y_{m_1}^{\left(l_1\right) } }
C^{\left(k\right)}_{m_1-m_2} 
\Ket{Y_{m_2}^{\left(l_2\right)}}
\end{align}
are angular integrals of spherical harmonics $Y_{m}^{\left(l\right) }$ and renormalized spherical harmonics $C^{\left(k\right)}_{m} = \sqrt{\frac{4 \pi}{2 k +1}} Y_{m}^{\left(k\right)}$.
The Slater integrals $F^{(2)}$ and $F^{(4)}$ are related to the multipole interaction between two $d$ electrons and responsible for the multiplet splitting between the different levels. They can be approximated by $J_H$ albeit at the loss of the experimentally observed multiplet structure. $F^{(0)}$ is the spherical averaged Coulomb interaction, i.e. the monopole part of the interaction. $F^{(0)}$ is related to $U$ by:
\begin{equation}
F^{(0)} = U + \frac{2}{63}(F^{(2)} + F^{(4)}).
\end{equation}

\subsection*{\texorpdfstring{Coulomb repulsion between a Ni $2p$ and Ni $3d$ electron - $H_{U}^{pd}$}{}}

The onsite interaction between the Ni $2p$ and Ni $3d$ electrons is given as:
\begin{align}
H_{U}^{pd} &= \sum_{\tau_1,\tau_2,\tau_3,\tau_4} 2 U_{\tau_1,\tau_2,\tau_3,\tau_4}^G \bm{d}^{\dag}_{\tau_1} \bm{p}^{\dag}_{\tau_2} \bm{p}^{\phantom{\dag}}_{\tau_3} \bm{d}^{\phantom{\dag}}_{\tau_4} + 2 U_{\tau_1,\tau_2,\tau_3,\tau_4}^F \bm{d}^{\dag}_{\tau_1} \bm{p}^{\dag}_{\tau_2} \bm{d}^{\phantom{\dag}}_{\tau_3} \bm{p}^{\phantom{\dag}}_{\tau_4},
\end{align}
with 
\begin{align}
 U_{\tau_1,\tau_2,\tau_3,\tau_4}^F = -\frac{1}{2} \delta_{\sigma_1,\sigma_3} \delta_{\sigma_2,\sigma_4}
 \sum_{k=0,2} 
 c^{\left(k\right)}\left[l_1=2,m_1;l_3=2,m_3\right]
 c^{\left(k\right)}\left[l_4=1,m_4;l_2=1,m_2\right]
 \times F^{\left(k\right)}_{pd},
\end{align}
and 
\begin{align}
 U_{\tau_1,\tau_2,\tau_3,\tau_4}^G = -\frac{1}{2} \delta_{\sigma_1,\sigma_3} \delta_{\sigma_2,\sigma_4}
 \sum_{k=1,3} 
 c^{\left(k\right)}\left[l_1=2,m_1;l_3=1,m_3\right]
 c^{\left(k\right)}\left[l_4=2,m_4;l_2=1,m_2\right]
 \times G^{\left(k\right)}_{pd}.
\end{align}
The monopole part of the $p$-$d$ interaction ($F^{(0)}_{pd}$) is related to $Q = U_{pd}$ by:
\begin{equation}
F^{(0)}_{pd} = U_{pd} + \frac{1}{15}G^{1}_{pd} + \frac{3}{70}G^{3}_{pd}
\end{equation}

\vspace*{10pt}

\subsection*{\texorpdfstring{Ni $3d$ and $2p$ spin-orbit coupling - $H_{\bm{l} \cdot \bm{s}}^{d}$ and $H_{\bm{l} \cdot \bm{s}}^{p}$}{}}

The spin-orbit coupling interaction within either the $2p$ or $3d$ shell of Ni is given as:
\begin{align}
H_{\bm{l \cdot s}} = \xi\sum_{i} \bm{l}_i \cdot \bm{s}_i =
\xi\sum_{m=-l}^{m=l} \sum_{\sigma} m\sigma \bm{a}^{\dagger}_{m\sigma} \bm{a}_{m\sigma} +
\frac{\xi}{2}\sum_{m=-l}^{m=l-1} \sqrt{l+m+1}\sqrt{l-m}
\left( 
\bm{a}^{\dagger}_{m+1,\downarrow} \bm{a}_{m,\uparrow} +  
\bm{a}^{\dagger}_{m,\uparrow} \bm{a}_{m+1,\downarrow}
\right)
\end{align}
where here $m$ is the orbital index, $l$ the angular momentum, i.e. either $p$ or $d$, $\sigma$ is the spin index, and $\xi$ the coupling constant. The operator $\bm{a}^{\dagger}$ corresponds to $\bm{d}^{\dagger}$ or $\bm{p}^{\dagger}$ for the valence $3d$ or core $p$ shells respectively.

\vspace*{10pt}

\subsection*{Hopping Parameter Definitions}

Both the inter- and intra-cluster hybridization interactions are affected by the breathing distortion. This dependence is contained in the hopping integrals $V_{e_{g}}$ and $V_{t_{2g}}$ of equations \ref{Eqn:Hhyb} and \ref{Eqn:Hmix}. In terms of the non-breathing values $V_{e_{g},0}$ and $V_{t_{2g},0}$, the hybridization parameters are defined using Harrison's rules \cite{Harrison1983S}, as
\begin{align}
V_{e_{g}} &= V_{e_{g},0} \left( 1 + \frac{\delta d}{d_0} \right)^{-4} \\
V_{t_{2g}} &= V_{t_{2g},0} \left( 1 + \frac{\delta d}{d_0} \right)^{-4}
\end{align}
where $\delta d$ is the bond length displacement as defined above. The values used for $V_{e_{g},0}$ and $V_{t_{2g},0}$ are provided in the section below. 

\clearpage

\section*{Parameter Values}

Parameter values enter into our model in the form of Coulomb interactions, on-site energies, spin orbit interactions, and hopping integrals. The values for these parameters have been quite well established over several decades of core level spectroscopy and other techniques, and here we do not deviate in any way from standard values (discussions of typical values can be found in references \onlinecite{degroot2008S, Zaanen_NiPES_PRB_1986S, Zaanen_ZSA_PRL_1985S, Haverkort_Wannier_PRB_2010S, Ballhausen1968S}, among many others). For the monopole Coulomb interaction parameters we use $U_{dd} = 6$ eV and $U_{pd} = 7$ eV. For the charge transfer energy we use $\Delta = -0.5$ eV (see the following paragraph for a discussion of the exact definition of $\Delta$). For the non-breathing-distorted values of on-site energies we use $10Dq = 0.95$ eV and $T_{pp} = 0.75$ eV. For non-breathing-distorted intra-cluster hopping integrals we use $V_{e_{g},0} = 3.0$ eV and $V_{t_{2g},0} = 1.74$ eV. For inter-cluster hopping, we find best agreement with experiment using $V_{I} = \sqrt{x} = 0.35$. Spin orbit interaction parameters are taken as the atomic values for Ni $3d^7$, $\xi_{2p} = 11.506$ eV  and $\xi_{3d} = 91$ meV. Finally, the multipole Coulomb interaction parameters are taken as 80\% of their atomic Hartree-Fock values for Ni $3d^7$ \cite{Cowan1981S}: $F^2_{dd} = 10.622$, $F^4_{dd} = 6.636$, $F^2_{pd} = 6.680$, $G^1_{pd} = 5.066$, and $G^3_{pd} = 2.882$, all expressed in units of electron volts.

There exist several different ways to define the charge transfer energy $\Delta$. As detailed in the preceding sections, in our model $\Delta$ contributes to the shell average energies $\epsilon_d$, $\epsilon_L$, and $\epsilon_p$. Thus, our value of $\Delta = -0.5$ eV sets the energy of the $d^8\underline{L}$ configuration 0.5 eV lower than the $d^7$ configuration \emph{before} the inclusion of multiplet, spin-orbit, crystal field, and ligand-ligand hopping terms. These terms all modify the energy separation between the lowest $d^7$ and $d^8\underline{L}$ states, thus affecting the effective charge transfer energy.  For example, the ligand-ligand hopping shifts the ligand $e_g$ orbitals up by $T_{pp}$ and shifts the ligand $t_{2g}$ orbitals down by $T_{pp}$, which ends up shifting the $d^8\underline{L}$ configuration lower by $T_{pp}$ with respect to the $d^7$ configuration, thus decreasing the effective delta further. Once all of the on-site energy effects are considered, we find that the lowest $d^8\underline{L}$ eigenstate is 0.6 eV lower than the lowest $d^7$ eigenstate (before the inclusion of hybridization), thus giving an effective charge transfer energy of $-0.6$ eV.

\section*{Spectroscopy Simulations}

After finding the ground state wavefunction of our model, we simulate the core XAS and MCD spectra using a Lanczos-based Green's function method \cite{Dagotto_RMP1994S, Haverkort_Wannier_PRB_2010S, Haverkort_DMFTXAS_EPL2014S}. For XAS we calculate the isotropic signal via the sum of spectra using $z$, left circular, and right circular polarized dipole transition operators. The MCD response (i.e. the fundamental MCD spectrum, or $f^{\left(1\right)}$ component of the atomic scattering tensor) is the difference between left and right circular polarized spectral functions.

The resonant magnetic diffraction signal of the nickelates has been intensively studied of late, with experiments probing the collinearity of moments via the azimuthal dependence of the diffraction signal \cite{Frano_OrbControl_PRL_2013S}. For this work we are interested in the energy dependence of the RMD signal, so for simplicity we assume a collinear arrangement of spins with a structure factor of 
\begin{align}
S\left[\bm{q}=\left(\sfrac{1}{4},\sfrac{1}{4},\sfrac{1}{4}\right)\right] = 
1\cdot f^{\left(1\right)}_{A1} + i\cdot f^{\left(1\right)}_{B1} - 1\cdot f^{\left(1\right)}_{A2} - i\cdot f^{\left(1\right)}_{B2}
\end{align} 
where $f^{\left(1\right)}$ denote the complex valued, energy dependent, magnetic circular dichroic form factors and $A1$ and $A2$ refer to short bond sites in different antiferromagnetic planes (similar for long bond sites $B$). Given an antiferromagnetic alignment of $A1$ and $A2$ and thus a negation of the form factor ($f^{\left(1\right)}_{A1} = -f^{\left(1\right)}_{A2} \equiv f^{\left(1\right)}_{A}$, and similar for $B$), our structure factor simplifies to
\begin{align}
S\left[\bm{q}=\left(\sfrac{1}{4},\sfrac{1}{4},\sfrac{1}{4}\right)\right] = 
2 \left( f^{\left(1\right)}_{A} + i\cdot f^{\left(1\right)}_{B} \right)
\end{align} 
and, neglecting polarization effects for simplicity, the spectral intensity is therefore
\begin{align}
\label{Eqn:RMD}
I\left(\omega\right) \propto \left| f^{\left(1\right)}_{A}\left(\omega\right) + i f^{\left(1\right)}_{B}\left(\omega\right) \right|^2
\end{align}
where we have restored the dependence on energy, $\omega$. Note that this expression does assume a particular domain structure, as $S_A \leq S_B$ in our model, i.e. the ordering in Eqn. \ref{Eqn:RMD} is (~$\SpinUp[6pt]~\SpinUp[10pt]~\SpinDn[6pt]~\SpinDn[10pt]~$).  One can assume the alternate ordering (~$\SpinUp[10pt]~\SpinUp[6pt]~\SpinDn[10pt]~\SpinDn[6pt]~$) with $f^{\left(1\right)}_{A}$ and $f^{\left(1\right)}_{B}$ switched in Eqn. \ref{Eqn:RMD} and obtain a slightly different energy dependence of the magnetic scattering. The spectra of both arrangements are shown along with the XAS for $\delta d = 0.03$\AA~in Fig. \ref{Fig:SRMD} below. While both spectra are qualitatively similar, having the magnetic scattering intensity strongest at the energy of the first XAS peak and a small shoulder at higher energy, there are differences in overall intensity and in the fine details of the peaks. In the manuscript, we plot the average of these two responses.

\begin{figure}[t]
\includegraphics[width=4in]{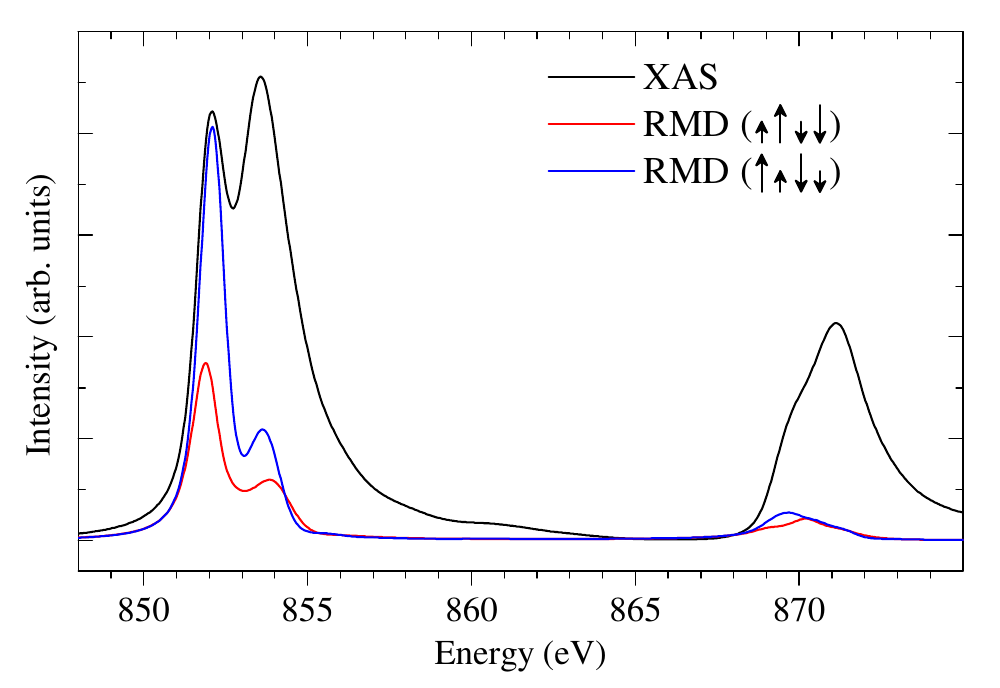}
\caption{Comparison of resonant magnetic diffraction at a breathing distortion of $\delta d = 0.03$\AA ~for the two different spin arrangements shown. }
\label{Fig:SRMD}
\end{figure}

\end{document}